\documentclass[acmsmall]{acmart}

\setcopyright{cc}
\setcctype{by}
\acmJournal{PACMMOD}
\acmYear{2026} \acmVolume{4} \acmNumber{6 (SIGMOD)} \acmArticle{337}
\acmMonth{12} \acmDOI{10.1145/3856332}

\usepackage{amsmath,amssymb,amsfonts}
\usepackage{graphicx}
\usepackage{textcomp}
\usepackage{color}
\usepackage{hyperref}
\usepackage{adjustbox}
\usepackage{makecell}
\usepackage{microtype}
\usepackage{multirow}
\usepackage{threeparttable}
\usepackage{enumitem}
\usepackage{booktabs}
\usepackage{bbding}
\usepackage{colortbl}
\usepackage{xcolor}
\usepackage{setspace}
\usepackage{pifont}
\usepackage{caption}
\usepackage{subcaption}
\usepackage{url}
\usepackage{fontawesome}
\usepackage{listings}
\usepackage[ruled,linesnumbered,vlined]{algorithm2e}
\usepackage[noend]{algpseudocode}
\usepackage[most]{tcolorbox}
\usepackage{float}
\newcommand{\eg}{\emph{e.g.,}}
\newcommand{\ie}{\emph{i.e.,}}

\newcommand{\polygon}{\textsc{Polygon}}
\newcommand{\verieql}{VeriEQL}
\newcommand{\parseval}{ParSEval}

\newcommand{\codeIn}[1]{\texttt{\small\detokenize{#1}}}
\newcommand{\tool}{SQLWitness}
\newcommand{\bird}{\textsc{Bird}}
\newcommand{\parrot}{\textsc{Parrot}}
\newcommand{\sede}{\textsc{Sede}}

\lstdefinestyle{tsql}{
  basicstyle=\ttfamily\footnotesize,
  keywordstyle=\bfseries,
  morekeywords={SELECT,FROM,WHERE,INNER,LEFT,RIGHT,FULL,CROSS,GROUP,BY,HAVING,DISTINCT,JOIN, AS, AND, "<"},                               
  emphstyle=\color{purple}\bfseries,                    
  showstringspaces=false,
  tabsize=2,
  captionpos=t,
  frame=none,
  rulecolor=\color{black},
  escapeinside={(*@}{@*)},
  moredelim=**[is][\color{red}]{@}{@},
  xleftmargin=-2em,
  belowskip=-1em,
}

\lstdefinestyle{sql}{
  basicstyle=\ttfamily\footnotesize,
  keywordstyle=\bfseries,
  morekeywords={SELECT,FROM,WHERE,INNER,LEFT,RIGHT,FULL,CROSS,GROUP,BY,HAVING,DISTINCT,JOIN},
  emph={INNER,LEFT,RIGHT,FULL,CROSS,DISTINCT, ALL, AND, OR, ORDER, AS, ON, CREATE, TABLE, INSERT, INTO, VALUES, INTEGER, LIMIT, VARCHAR},
  emphstyle=\bfseries,
literate={<}{{{\textbf{<}}}}1
       {/}{{{\textbf{/}}}}1,
  showstringspaces=false,
  numbers=left,
  numberstyle=\footnotesize,
  stepnumber=1,
  numbersep=10pt,
  tabsize=2,
  captionpos=t,
  frame=none,
  rulecolor=\color{black},
  escapeinside={(*@}{@*)},
  moredelim=**[is][\color{red}]{@}{@},
  moredelim=**[is][\color{gray}]{~}{~},
  xleftmargin=1.5em,
  belowskip=-1em,
}

\begin{document}
\title{Efficient Dialect-Agnostic SQL Equivalence Refutation}

\author{Junwen An}
\orcid{0009-0003-1768-7443}
\affiliation{
  \institution{National University of Singapore}
  \country{Singapore}
}
\email{junwenan@u.nus.edu}

\author{Manuel Rigger}
\orcid{0000-0001-8303-2099}
\affiliation{
  \institution{National University of Singapore}
  \country{Singapore}
}
\email{rigger@nus.edu.sg}



\begin{abstract}
    
Identifying semantically inequivalent SQL queries is useful for many applications, such as evaluating Text-to-SQL and dialect translation systems. However, existing approaches often struggle with the diversity of SQL features and dialects, requiring significant manual effort and expertise to extend query feature support across Database Management Systems (DBMSs). To address these limitations, we present SQLWitness, an efficient, DBMS-agnostic query equivalence refutation engine. Our core idea is leveraging the DBMS execution engine to determine whether generated instances constitute counterexamples, thereby faithfully capturing the semantics of the SQL dialect and preventing false counterexamples. To efficiently generate test data, our approach explores a constrained data-generation space and uses an integrity-constraint-aware generation strategy. Furthermore, to guide time-efficient termination of this test-data generation process, we pair predicate-coverage signals with a statistical estimator. We extensively evaluated SQLWitness on six benchmark suites across four SQL dialects. SQLWitness refuted 5.34\% to 88.4\% more pairs than state-of-the-art baselines, and processed query pairs in 0.12 seconds on average, achieving a speedup from 2.1$\times$ to 16.4$\times$ over the baselines. Furthermore, in a case study, SQLWitness found 8.4\% inconsistent translations within a dialect translation benchmark. We believe SQLWitness could be broadly adopted for many use cases.

\end{abstract}
\maketitle

\section{Introduction}\label{sec:introduction}

\newcommand{\cmark}{\textcolor{green!60!black}{\ding{51}}}
\newcommand{\xmark}{\textcolor{red!70!black}{\ding{55}}}
\newcommand{\pmark}{\textcolor{orange!80!black}{$\sim$}}

\newcommand{\rot}[1]{\rotatebox{55}{#1}}

\begin{table}[]
  \centering
  \caption{%
    Comparison of representative tools. IC: Integrity Constraints; Gen W: Concrete Witness Generation.
    (\cmark~full support, \pmark~partial support, \xmark~no support).
   }
  \label{tab:tool-comparison}
  \small
  \setlength{\tabcolsep}{4pt}
  \begin{tabular}{l l cccc}
    \toprule
    \textbf{Tool} & \textbf{Approach} & \textbf{DBMS-Agnostic} & \textbf{IC.} & \textbf{Gen W.} & \textbf{Verify Eq.} \\
    \midrule
    \tool{}
      & Test-case
      & \cmark
      & \cmark
      & \cmark
      & \xmark
      \\
    ParSEVal~\cite{chen25Parseval}
      & Test-case
      & \pmark
      & \pmark
      & \cmark
      & \xmark
      \\
    VeriEQL~\cite{verieql24he}
      & Verification
      & \xmark
       & \cmark
      & \cmark
      & \pmark
      \\
    QED~\cite{wang24QED}
      & Verification
      & \xmark
      & \pmark
      & \xmark
      & \cmark
      \\
    \bottomrule
  \end{tabular}
\end{table}

Identifying when two SQL queries are \emph{semantically inequivalent} is valuable in many settings. Unlike proving equivalence, which is undecidable in general~\cite{chaudhuri94equiv, chauduri93optimization} and requires a formal model to verify that two queries produce identical results on \emph{all} possible inputs, inequivalence requires only a single concrete \emph{counterexample}, namely a database instance on which the two queries return different outputs (\ie{} a witness). Finding inequivalence witnesses is useful for evaluating Text-to-SQL systems~\cite{klopfenstein26spotit}, grading SQL assignments~\cite{chandra2019AutomatedGradingSQL}, and debugging queries~\cite{miao2019explaining}, where a counterexample exposes semantic differences between a ground-truth query and a candidate. In dialect translation~\cite{zhou2025cracking}, a counterexample between a source and translated query indicates an incorrect translation.


Table~\ref{tab:tool-comparison} summarizes representative existing works that identify semantically inequivalent queries, which can be generally classified into two categories based on their underlying approach. The first is verification-based, relying on formal verification by translating SQL queries into logical formulas and using automated solvers to prove query equivalence or construct counterexamples~\cite{verieql24he, chu17cosette, wang24QED}. However, existing verification-based tools involve tradeoffs in supported features. VeriEQL~\cite{verieql24he} supports witness generation and integrity constraints, but provides only \emph{bounded} equivalence guarantees, verifying equivalence up to a fixed database size. QED~\cite{wang24QED} can fully prove equivalence, but lacks witness generation and foreign key constraint support, limiting its applicability. The second category is test-case-based, where, rather than reasoning over logical formulas, these approaches generate test cases from the query structure or semantics to cover a range of query behaviors, such as varying predicate evaluation outcomes, and use execution to detect whether a counterexample exists~\cite{castelin2018searchbased, chen25Parseval, 2020spiderts}. While this latter class cannot formally prove equivalence, it can effectively determine the existence of a witness to inequivalence.

Two main challenges in equivalence refutation are the diversity of SQL features and the efficiency of the tools. A DBMS typically contains many dialect-specific SQL features, such as functions and data types, which require developers of query refutation and reasoning tools to manually implement support for such features. Across different DBMS vendors, this problem is further compounded, as varying SQL dialects can introduce semantic differences such as inconsistent data type conversion behaviors~\cite{elucidatingType25Ye} and incompatible dialect-specific features~\cite{zhong2024UnderstandingReusingTest}. There has been a sustained effort to push the boundaries of supported features in existing tools. For example, \parseval{}~\cite{chen25Parseval} applies symbolic analysis on query plans as a DBMS-agnostic intermediate representation while also supporting more features than previous tools, yet still fails on complex features such as string comparisons and window functions. More broadly, existing approaches, whether verification-based or test-case-based, typically operate on a fixed semantic model of queries and therefore might fail to distinguish semantic differences across DBMSs or require significant manual effort to support the semantics of each system. Beyond SQL feature support, efficiency remains a concern. Verification-based approaches in particular can struggle to scale, as reasoning over logical formulas becomes computationally expensive for complex queries, such as when multiple tables are joined~\cite{verieql24he}. Test-case-based approaches are generally more efficient, but when run until a timeout or until full branch coverage is achieved~\cite{2020spiderts, chen25Parseval}, they can still consume significant time.


In this paper, we propose \tool{}, a DBMS-agnostic and efficient query equivalence refutation engine that executes candidate queries against a DBMS, allowing each DBMS to interpret query semantics in its own dialect. \tool{} provides no guarantee of finding a counterexample for every inequivalent query pair, but every counterexample it reports is genuine, meaning that the result difference can be observed when the two queries are executed on the target DBMS. To facilitate database generation, we leverage \emph{fuzzing}~\cite{AFLplusplus-Woot20}, a biased randomized search technique, to efficiently explore query execution behaviors, eschewing explicitly modeling SQL features. We make three core technical contributions. First, \tool{} constructs a minimal data-generation space containing a set of values designed to help explore query execution behaviors, which we term \emph{candidate pools}. These include initial values based on three-valued logic (3VL) to explore structural behaviors (\eg{} table joins), alongside literal values extracted from the queries to satisfy predicates. Second, \tool{} uses an integrity-constraint-aware sampling strategy over candidate pools to generate valid database instances. Third, \tool{} measures fuzzing progress by running predicate coverage queries on the instances and employing a statistical estimator to terminate fuzzing once exploration saturates.

We extensively evaluated \tool{} on six benchmark suites comprising 26,665 query pairs across four dialects. \tool{} supported all query pairs and outperformed state-of-the-art approaches across all benchmarks, with 5.34\% to 88.4\% more refuted query pairs. Furthermore, \tool{} proved highly efficient, as it processed a query pair in 0.12 seconds on average, achieving a 2.1$\times$ to 16.4$\times$ speedup over all existing baselines. Additionally, we conducted a case study using \tool{} to evaluate the semantic equivalence of queries in dialect translation benchmarks. We found that 19.6\% of the translated queries contained execution errors, and 8.4\% failed to account for semantic differences.

We believe that \tool{} can be broadly adopted for many use cases due to its DBMS-agnostic nature. For example, it can be used to evaluate dialect translation tools~\cite{zhou2025cracking} and Text-to-SQL systems, where benchmarks are becoming increasingly diverse in their dialects~\cite{pourreza2024SQLGENBridgingDialect}. Additionally, since it is highly efficient, users can pair \tool{} with constraint-solver-based tools, running \tool{} for fast refutation before invoking a more heavyweight solver without causing significant performance overhead. Our artifact is publicly available.\footnote{We released a permanent artifact at https://doi.org/10.5281/zenodo.23396996 as well as the up-to-date version at https://github.com/wanteatfruit/SQLWitness.}

To summarize, we make the following contributions:
\begin{itemize}[leftmargin=*]
    \item We propose the novel idea of query-aware fuzzing to refute SQL query equivalence in a DBMS-agnostic manner.
    \item We present \tool{}, a query equivalence refutation engine that implements this idea through candidate pools, a constraint-aware data generation strategy, and a predicate coverage mechanism with a statistical termination estimator.
    \item We extensively evaluated \tool{} on over 26,000 query pairs across six benchmarks and a real-world case study.
\end{itemize}
\section{Background}\label{sec:background}
In this section, we present the background and formal problem definition of our work.

\paragraph{SQL Semantics and Dialects}
SQL is a declarative language for querying and manipulating data in Relational Database Management Systems (RDBMSs). SQL fundamentally relies on three-valued logic (3VL), where predicates evaluate to \codeIn{TRUE}, \codeIn{FALSE}, or \codeIn{NULL}~\cite{codd793vl}.

While the core relational algebra underlying SQL relies on well-established formal semantics, the SQL standard itself~\cite{ansi-sql} leaves many query behaviors implementation-dependent. Consequently, DBMSs implement their own distinct SQL variants, known as \emph{SQL dialects}~\cite{critiqueOfModern24thomas}. We use the terms ``DBMS'' and ``dialect'' interchangeably, as each DBMS inherently defines its own dialect. 

SQL dialects frequently diverge from both the standard and one another in their syntax, semantics, and supported feature sets. For example, the \codeIn{||} operator represents a logical \codeIn{OR} in MySQL, but string concatenation in PostgreSQL and SQLite. Even when dialects share an operator, precedence rules can conflict: the expression \codeIn{3 * 2 || `1'} evaluates to \codeIn{TRUE} in MySQL, \codeIn{63} in SQLite, and \codeIn{61} in PostgreSQL. Beyond conflicting interpretations of shared features, dialects frequently provide unique features. For instance, PostgreSQL provides the \codeIn{ILIKE} operator for case-insensitive pattern matching, whereas systems like MySQL and SQLite lack this specific operator and handle case-sensitivity through default collations. Furthermore, Ye et al.~\cite{elucidatingType25Ye} demonstrated that type conversion behaviors across dialects also exhibit widespread inconsistencies. Ultimately, these DBMS-specific behaviors make static semantic encodings inherently impractical, as significant manual effort is needed to formally encode and maintain the rules for each dialect.

\paragraph{Problem Definition}
Given two SQL queries $Q_1$ and $Q_2$ in dialects $D_1$ and $D_2$, a database schema
$S$, and a set of integrity constraints $C$, our goal is to find a concrete database
instance $d$ that witnesses their inequivalence. $Q^D(d)$ denotes the result of \emph{executing} query $Q$ in dialect $D$ on instance $d$ against an actual DBMS. We adopt result-based equivalence, where two queries are considered inequivalent if their execution results differ. This is motivated by real-world use cases like debugging~\cite{chi20debugging} and dialect translation~\cite{zhou2025cracking}, where observable results are important. In this setting, we also consider two ambiguous, but equivalent queries that produce different results as non-equivalent. Formally, we say $Q_1$ and $Q_2$
are \emph{semantically equivalent} when $\forall d \text{ s.t. } d \models C,\,
Q_1^{D_1}(d) = Q_2^{D_2}(d)$, and \emph{semantically inequivalent} when $\exists d
\text{ s.t. } d \models C,\, Q_1^{D_1}(d) \neq Q_2^{D_2}(d)$. A \emph{genuine
counterexample} is an instance $d$ satisfying two conditions: it adheres to all
integrity constraints $C$, and executing both queries on it produces different results,
\ie{} $Q_1^{D_1}(d) \neq Q_2^{D_2}(d)$.

This problem arises most directly within a single DBMS ($D_1 = D_2$), where a genuine counterexample is evidence of inequivalence. The majority of our evaluation targets the single-DBMS setting. The problem can also be generalized to queries across different DBMSs, which we discuss in the context of a dialect translation use case in Section~\ref{sec:usecase}.

An effective tool should minimize both false positives and false negatives. A false positive, or spurious counterexample, is an instance $d$ reported by a tool
for which the queries return the same result on the actual DBMS, \ie{}
$Q_1^{D_1}(d) = Q_2^{D_2}(d)$. When comparing the results, if neither query has an \codeIn{``ORDER BY''} clause, we verify only that the result sets contain the same values, ignoring their order. If at least one query has an \codeIn{``ORDER BY''} clause, we compare both the values and ordering.  A \emph{false negative} occurs when a tool fails to find a counterexample for a genuinely inequivalent pair. Notably, since SQL query equivalence is undecidable in general~\cite{chauduri93optimization, chaudhuri94equiv}, false negatives are generally unavoidable for any terminating tool.



\paragraph{Fuzz Testing}

Fuzz testing, or fuzzing, is a widely adopted dynamic analysis technique that finds bugs in software by generating inputs aiming to trigger incorrect states. \emph{Black-box} fuzzers~\cite{miller1990fuzz} treat the program as opaque and generate inputs from a specification or through random mutation, without using execution feedback. \emph{White-box} fuzzers~\cite{godefroid2008sage} rely on heavyweight program analysis, such as symbolic execution, to systematically explore execution paths; this can be precise, but often incurs a high run-time overhead. \emph{Grey-box} fuzzers, such as AFL++~\cite{AFLplusplus-Woot20}, instead use approximate run-time signals, typically code coverage, to guide input generation toward unexplored behaviors. A fundamental component of a fuzzer is its \emph{input space}, which defines the seed values that the fuzzer could explore. To measure progress, coverage-guided fuzzers rely on a \emph{coverage metric} (\eg{} branch coverage) to quantify how thoroughly the program's behavior has been explored. Finally, \emph{stoppage criteria} determine when to terminate the fuzzing campaign, typically when the probability of discovering new coverage falls below a specific threshold~\cite{bohme2018stads, estimatResidual21Bohme}. Building on these concepts, \tool{} defines its \emph{input space} as data values and table sizes (\ie{} cardinality), uses \emph{predicate coverage} to approximate query behavior exploration, and applies a statistical estimator as its \emph{stoppage criterion}.

\begin{figure*}
    \centering
    \includegraphics[width=\linewidth]{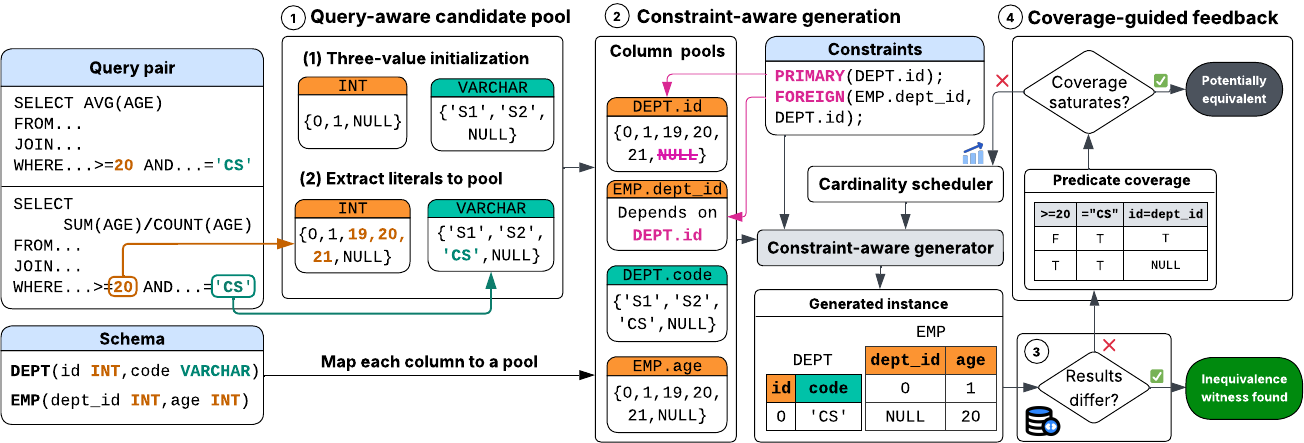}
    \caption{Illustrative example of \tool{}.}
    \Description{Illustration}
    \label{fig:walkthrough}
\end{figure*}

\begin{lstlisting}[style=sql, label={lst:schema}, caption={Example schema and data}, float]
CREATE TABLE DEPT(id INTEGER PRIMARY KEY, code VARCHAR);
CREATE TABLE EMP(dept_id INTEGER REFERENCES DEPT(id), age INTEGER);
INSERT INTO DEPT VALUES (0, 'CS');
INSERT INTO EMP VALUES (0, 20), (0, 21)
\end{lstlisting}



\begin{lstlisting}[style=sql, label={lst:avg}, caption={Average function vs. integer division}, float]
SELECT @AVG(age)@ FROM EMP
JOIN DEPT ON DEPT.id = EMP.dept_id
WHERE EMP.age >= 20 AND DEPT.code = 'CS';
~-- SQLite/Postgres (20.5); MySQL (20.5); SQL Server (20)~

SELECT @SUM(age)/COUNT(age)@ FROM EMP
JOIN DEPT ON DEPT.id = EMP.dept_id
WHERE EMP.age >= 20 AND DEPT.code = 'CS';
~-- SQLite/Postgres (20); MySQL (20.5); SQL Server (20)~
\end{lstlisting}

\section{Motivating Example}

We use the schema in Listing~\ref{lst:schema} together with the query pair in Listing~\ref{lst:avg} to illustrate the challenges of query refutation faced by existing tools and to show how \tool{} constructs data instances to expose such inequivalences. The schema contains two tables, \codeIn{DEPT} and \codeIn{EMP}, each with two columns. The \codeIn{id} column in \codeIn{DEPT} is a \emph{primary key}, meaning it is \codeIn{NOT NULL} and \codeIn{UNIQUE} across all rows. The \codeIn{dept_id} column in \codeIn{EMP} is a \emph{foreign key} referencing \codeIn{DEPT.id}, requiring that every non-null value in \codeIn{dept_id} exist in \codeIn{DEPT.id}. For the query pair, we highlight their difference in red. Both queries in Listing~\ref{lst:avg} retrieve the average age of employees aged 20 or older who belong to the \texttt{`CS'} department. They differ only in how the average is computed, with one using the built-in \codeIn{AVG} function and the other computing the ratio of \codeIn{SUM} to \codeIn{COUNT} directly.

Although the two queries appear logically equivalent, their results can diverge at run time because the \codeIn{AVG} function and the \codeIn{/} operator may carry different type-conversion semantics depending on the DBMS~\cite{elucidatingType25Ye}. On SQLite, \codeIn{AVG} promotes its result to a floating-point value and returns 20.5, whereas integer division truncates and returns 20. MySQL implicitly widens integer division to a floating-point result, so both queries return 20.5. SQL Server, by contrast, preserves integer types throughout and returns 20 for both expressions. These behavioral differences highlight a fundamental challenge for existing query refutation tools, where encoding a fixed formal semantics may fail to capture the diverging interpretations adopted by individual DBMSs.

Existing approaches refute equivalence by encoding a specific SQL semantics, but this encoding cannot capture the actual run-time behavior across all DBMSs. For example, as a result, \verieql{} classifies this pair as equivalent, which is incorrect when the pair is executed on SQLite or PostgreSQL. Although existing approaches could, in principle, be extended to cover more DBMS-specific behaviors, doing so would require substantial domain expertise and continuous maintenance effort as DBMSs evolve.


The core of our approach is to use a biased randomized search technique that executes the queries over dynamically generated data on the DBMS, thus accurately capturing its behavior. Rather than modeling query semantics, we model the space of database instances explored by the fuzzer. Our key insight is that a small number of concrete values is often sufficient to expose many divergent behaviors while keeping the input space small for efficient termination. Accordingly, we perform only minimal syntactic analysis of the query pair to populate the input space, delegating all semantic interpretation to the underlying DBMS during execution. This allows \tool{} to support all SQL features and remain DBMS-agnostic. In contrast, tools such as \verieql{} and \parseval{} require explicit semantic modeling of queries or operators, which limits their SQL feature coverage and efficiency on complex queries.


Figure~\ref{fig:walkthrough} illustrates the approach of \tool{}, which proceeds in four steps. First, it constructs a small, per data type candidate pool that serves as the fuzzer's input space (\ding{172}). For each type, we initialize each pool with two arbitrarily distinct concrete values and a \codeIn{NULL} value. For example, we initialize the pool for \codeIn{INT} as \codeIn{{0, 1, NULL}}. These three values would be sufficient to explore the fundamental comparison and join behaviors of SQL under its ternary logic, where the two concrete values cover equality and inequality comparisons, and \codeIn{NULL}-value-specific behaviors. For instance, a single \codeIn{NULL} value in a joined column is sufficient to distinguish a \codeIn{LEFT JOIN} from an \codeIn{INNER JOIN}, without enumerating the full input space that might cause path explosion in formal approaches. However, three arbitrary values are insufficient when queries contain literal-guarded predicates. In our example, both queries filter on \codeIn{age >= 20}, so the initial pool never exercises this path. To address this, we apply \emph{literal-based seeding}, where we extract literal values directly from the query and add them, along with their boundaries, to the candidate pool of the matching type. For \codeIn{INT}, we add the literal and its neighbors (\eg{} \codeIn{{19, 20, 21}}) to cover boundary conditions around comparison predicates, and for \codeIn{VARCHAR}, we add the literal itself (\eg{} \codeIn{`CS'}) since string comparisons are typically exact~\cite{szlang25snowflakeworkload}. 


Database instance generation proceeds by mapping each candidate pool to schema columns by data type, while enforcing integrity constraints (\ding{173}). For example, the \codeIn{INT} pool maps to all integer-typed columns (\codeIn{id}, \codeIn{dept_id}, and \codeIn{age}). Although certain values are only semantically relevant to specific columns, such as the boundary values around \codeIn{20} being meaningful only for the \codeIn{age} predicate, \tool{} deliberately avoids modeling such column-level relevance. Doing so would require understanding operator semantics and column roles in the query, reintroducing the feature-support limitations that \tool{} is designed to avoid. Instead, the small pool size allows useful values to be selected with high probability regardless. The column-level pools are then passed to the \emph{constraint-aware generator}, which first eliminates values that violate column constraints. For instance, the \codeIn{PRIMARY KEY} constraint on \codeIn{DEPT.id} removes \codeIn{NULL} from its pool. The \emph{cardinality scheduler} then determines how many rows to generate per table, starting small for efficiency and growing across iterations to explore a wider range of data combinations. We detail this process in Section~\ref{sec:constraint}. The generated instance is a constraint-satisfying database instance ready for execution, as shown in the ``Generated instance'' table in Figure~\ref{fig:walkthrough}. 


During fuzzing, a key challenge is determining when the query's behavior has been sufficiently explored to allow termination. In each iteration, \tool{} executes both queries on the generated database instance and compares their results (\ding{174}). Otherwise, \tool{} must evaluate its exploration progress to decide whether to continue (\ding{175}). Existing test-case-based approaches require precise semantic modeling to determine whether all relevant execution paths have been explored, such as the query plan coverage used by \parseval{}. For \tool{}, we instead propose \emph{approximate predicate coverage (APC)}, a coverage signal that approximates query behavior without the need for formal semantics modeling. Specifically, we extract the query pairs' predicates (\eg{} \codeIn{age >= 20}) and execute them on the generated data to directly obtain each predicate's evaluation result. APC evaluates these extracted predicates over the cross product of the referenced tables; thus, a multi-table predicate is evaluated on each corresponding row combination. We further discuss details in Section~\ref{sec:coverage}. In the example, the generated instance reveals two new results for \codeIn{age >= 20}, \codeIn{TRUE}, and \codeIn{FALSE}. Based on the results, \tool{} uses a statistical estimator to decide whether to continue fuzzing when new predicate outcomes are being discovered, or to terminate and declare the query pair as potentially equivalent when coverage has saturated. In the example, since one of the predicates produces new results, the estimator determines that continuing fuzzing could further explore the query. If the coverage feedback in subsequent iterations shows no new results from predicates, the cardinality scheduler would further increase the table size to encourage broader exploration. While the coverage signal provides no guarantee of finding a counterexample, the continued exploration would eventually produce an instance containing both \codeIn{age = 20} and \codeIn{age = 21}, which triggers a discrepancy in the queries' results and thus constitutes a counterexample.

\section{\tool{}}
\SetKwFor{While}{while}{do}{}
\SetKwInput{KwInput}{Input}
\SetKwInput{KwResult}{Result}
\SetKw{KwReturn}{return}
\SetAlCapFnt{\footnotesize}
\SetKwComment{Comment}{$\triangleright$\ }{}
\SetCommentSty{mycommfont}
\DontPrintSemicolon
\begin{algorithm}[t]
\footnotesize
\footnotesize\everymath{\footnotesize}
\setstretch{1.1}
\caption{\footnotesize SQLWitness}
\label{alg:sqlwitness}
\KwInput{\emph{$Q_1, Q_2, S, C, D_1, D_2, \delta$}}
\KwResult{$d\ \text{or}\ \textsc{PotentiallyEquivalent}$ \Comment*[r]{Witness or coverage saturated}}
\SetKwProg{Fn}{Function}{:}{}
\Fn{QueryAwareFuzzing}{
    \emph{pool} $\gets$ \emph{BuildCandidatePool}($Q_1, Q_2, S$) \Comment*[r]{Section 4.1}
    \emph{cov} $\gets \emptyset$;\ \emph{stagnate} $\gets$ \emph{0};\ \emph{bound} $\gets$ \emph{1} \;
    \While{$\hat{p}($stagnate$) \geq \delta$}{
        \emph{d} $\gets$ \emph{GenerateInstance}(\emph{pool, C, bound}) \Comment*[r]{Section 4.2}
        \uIf(\Comment*[f]{Execute and compare on DBMS}){Q$_1^{D_1}$(d) $\neq$ Q$_2^{D_2}$(d)}{
            \KwReturn{$d$} \;
        }
        \emph{cov}$_{\text{new}}$ $\gets$ \emph{ComputeCoverage}($Q_1, Q_2, d$) \Comment*[r]{Section 4.3}
        \uIf{{cov}$_{\text{new}}$ $\not\subseteq$ {cov}}{
            \emph{cov} $\gets$ \emph{cov} $\cup$ \emph{cov}$_{\text{new}}$;\ \emph{stagnate} $\gets$ \emph{0} \;
        }
        \Else{
            \emph{stagnate} $\gets$ \emph{stagnate} $+$ \emph{1};\ \emph{bound} $\gets$ \emph{bound} $+$ \emph{1}
        }
        \emph{bound} $\gets$ \emph{bound} $+$ \emph{1}\;
    }
    \KwReturn{\textsc{PotentiallyEquivalent}} \;
}
\end{algorithm}

We present \tool{}, an efficient DBMS-agnostic query equivalence refutation engine based on \emph{query-aware fuzzing}, which accurately models a dialect's SQL features by using the DBMS that implements it for evaluation. Algorithm~\ref{alg:sqlwitness} outlines the overall process. Its goal is to find a counterexample (line 7) by repeatedly generating database instances and executing both queries against each instance (lines 5--6); if the results differ, the instance is returned as a counterexample. Database instances are generated by sampling from the query-aware candidate pool under the integrity constraints $C$ (Section~\ref{sec:constraint}), with the row count bounded by an upper limit that increases each iteration to expose deeper execution states. The candidate pool is constructed once at initialization (line 2, Section~\ref{sec:pool}), before the main loop begins. Since the two queries may be equivalent, or \tool{} may fail to find a counterexample within a reasonable time budget, the fuzzer must also be able to determine when to terminate. To this end, \tool{} tracks coverage saturation across iterations (lines 8--13, Section~\ref{sec:coverage}). Specifically, a statistical estimator ($\hat{p}$) evaluates the stagnation count at the start of each loop (line 4) and triggers termination when coverage appears saturated, indicating that further generation is unlikely to find a counterexample (line 14).



\subsection{Query-Aware Candidate Pool}\label{sec:pool}
In this section, we detail \tool{}'s input space of data values, which we illustrate in Figure~\ref{fig:walkthrough} (\ding{172}). An effective input space must satisfy two competing requirements: (1) it should contain values capable of triggering divergent query behaviors, and (2) it should remain sufficiently small to maintain search efficiency. Guided by these requirements, \tool{} constructs an input space that retains only the essential seed values needed to target query execution paths that may expose differentiating results. To this end, we propose a heuristics-based \emph{query-aware candidate pool} approach, which models the input space of each data type as a discrete candidate set, seeded with three arbitrary initial values and augmented with literal values extracted from the queries. Formally, let $T$ be the set of all data types referenced in the schema and the candidate queries. For each type $t \in T$, we define its candidate pool as $P_t = I_t \cup L_t$, where $I_t$ is the set of initial seed values necessary to explore three-valued logic, and $L_t$ is the set of values of type $t$ extracted from the queries. The complete data value input space is thus defined as sets $\mathcal{P} = \{P_t \mid t \in T\}$. By sampling exclusively from these targeted pools rather than the entire domain of valid values, the fuzzer uses a drastically reduced input space and accelerates the exploration of query behaviors. Since candidate-pool construction is heuristic, it provides no guarantee that the resulting input space is sufficient to trigger bugs or that it is minimal. Still, we show in the evaluation that these heuristics are highly effective in practice. We subsequently detail both initialization steps.


\paragraph{Three-value-based initialization}

We initialize each column's candidate pool with three values. This initialization strategy is motivated by SQL's three-valued logic~\cite{codd793vl} and the prevalence of equi-join predicates in practice~\cite{schu16equijoins}. Formally, for a given data type $t$, we define its initial pool as $I_t = \{v_{1,t}, v_{2,t}, \codeIn{NULL}\}$, where $v_{1,t}$ and $v_{2,t}$ are two arbitrarily chosen, distinct concrete values of type $t$. For the initial pool, the specific values matter only in that they must be distinct, as predicates involving literal constants are handled separately by the seeding strategy described next. For example, while in Figure~\ref{fig:walkthrough} we chose $I_{\text{INT}} = \{0, 1, \codeIn{NULL}\}$ as the initial values, one could also use $\{-10, 10, \codeIn{NULL}\}$ and the same counterexample would still be found. The three initial values already provide a foundation to trigger different predicate evaluation paths. Specifically, the two distinct concrete values enable testing equality when values match (\codeIn{TRUE}) and inequality when values differ (\codeIn{FALSE}), while the \codeIn{NULL} value enables testing of \codeIn{NULL}-related behaviors. 


\paragraph{Literal-based seeding} We complement the default candidate pool with literal values extracted from the query pair, ensuring the fuzzer can explore behaviors guarded by specific predicates. Without this, a default pool of $I_{\text{INT}} = \{0, 1, \codeIn{NULL}\}$ cannot trigger a counterexample due to the predicate \codeIn{age >= 20} in Figure~\ref{fig:walkthrough}, as none of the candidate values can explore the query behavior where the predicate evaluates to \codeIn{TRUE}. To mitigate this, we adopt an effective heuristic based on Wu \emph{et al.}'s insight regarding fuzzing exploration strategies~\cite{wu25Tumbling}, which demonstrates that injecting a literal from the program under test into the fuzzer's input space significantly enhances program exploration. Consequently, we propose \emph{literal-based seeding}. Formally, let $V_t$ be the set of literal values of type $t$ extracted from the query pair. For each literal $v \in V_t$, we define a boundary function $B(v)$ that generates $v$ and its immediate semantic neighbors. The literal-seeded set is then defined as $L_t = \bigcup_{v \in V_t} B(v)$. For instance, a numeric literal $v$ contributes the set $B(v) = \{v-1, v, v+1\}$. Including these immediate boundary values is motivated by boundary value analysis, a well-established principle in software testing that targets values at and around boundary points to expose behavioral differences~\cite{myers2011art, fu25boundaryfunction}. For the predicate \codeIn{age >= 20}, we extract $20 \in V_{\text{INT}}$ and add $B(20) = \{19, 20, 21\}$ to the \codeIn{INT} pool. By generating values such as $20$ or $21$, the fuzzer satisfies the predicate and exposes a counterexample that is otherwise unreachable. Boundary rules can also be applied to any other data types. For example, temporal literals (\eg{} \codeIn{DATE}) are offset by plus or minus one unit, while string literals (\eg{} \codeIn{VARCHAR}) are seeded exactly as extracted, since text comparisons in real-world workloads typically involve equality predicates rather than ranges~\cite{szlang25snowflakeworkload}.

\paragraph{Grouping and aggregation}
\tool{} applies the same three-value-based initialization to grouping keys, allowing generated rows to be partitioned differently and thereby expose differences in \codeIn{GROUP BY} behavior. \tool{} avoids using aggregate-specific heuristics, as rules for individual functions may fail to generalize to dialect-specific aggregates. Rather, \tool{} relies on the target DBMS to evaluate aggregate semantics.

\subsection{Constraint-aware Generation}\label{sec:constraint}
With the candidate pools constructed in Section~\ref{sec:pool}, we now describe the constraint-aware data generator (see Figure~\ref{fig:walkthrough} (\ding{173})), which populates a concrete database instance by sampling from these pools while satisfying integrity constraints. Integrity constraints are widely used in schemas~\cite{till24schemapile}, making this support essential. Generating instances that violate integrity constraints would prevent their insertion into the database~\cite{verieql24he}. 

\tool{} models the semantics \codeIn{PRIMARY KEY}, \codeIn{FOREIGN KEY}, \codeIn{NOT NULL}, \codeIn{UNIQUE}, and simple \codeIn{CHECK} constraints (\eg{} column comparisons) during generation. \codeIn{CHECK} constraints involving DBMS-specific operators are difficult to encode without dialect-specific knowledge. To ensure soundness in all cases, \tool{} additionally embeds all integrity constraints in the \codeIn{CREATE TABLE} statements, so the DBMS validates every inserted row at run time; any instance the DBMS accepts is guaranteed to satisfy its actual constraint semantics. \tool{} satisfies integrity constraints through a staged generation procedure, namely (1) table ordering, (2) per-column candidate pool construction, and (3) value generation.

\paragraph{Table ordering}
First, \tool{} determines table metadata, including inter-table dependencies and valid cardinalities. Tables are ordered \emph{topologically} by their foreign key relationships.\footnote{In the rare case of cyclic foreign key dependencies, \tool{} breaks the cycle at a nullable foreign key column and fills it with \codeIn{NULL}.} In Figure~\ref{fig:walkthrough}, \codeIn{EMP.dept_id} depends on \codeIn{DEPT.id}, so \codeIn{DEPT} is generated first, allowing \codeIn{dept_id} to reference its values. Each table is assigned a row count drawn uniformly at random between 0 and an upper bound (detailed in Section~\ref{sec:coverage}). One special case may cause \tool{} to adjust this assignment. A table with a \codeIn{UNIQUE FOREIGN KEY} column cannot exceed the row count of its referenced parent, so such child tables' cardinalities are capped to their parents' cardinalities.

\paragraph{Per-column candidate pool}
Second, \tool{} initializes a per-column candidate pool by mapping each column to its type-based pool (\eg{} \codeIn{VARCHAR} or \codeIn{INT}) and filtering out values that violate constraints. In Figure~\ref{fig:walkthrough}, \codeIn{DEPT.code} maps to the \codeIn{VARCHAR} pool, while the remaining \codeIn{INT} columns map to the \codeIn{INT} pool. Values failing a constraint are then removed from the pool. For instance, \codeIn{NULL} is removed from the pool of \codeIn{DEPT.id} column, which is a \codeIn{PRIMARY KEY}. Similarly, values failing a \codeIn{CHECK} constraint are excluded. For example, if \codeIn{EMP.age} has a \codeIn{CHECK} constraint \codeIn{age > 18}, the values in \codeIn{{0, 1}} are removed from its pool. To preserve boundary coverage, \tool{} re-inserts the value closest to the boundary that still satisfies the constraint (\eg{} the smallest satisfying integer for a \codeIn{`>'} constraint) into the pool, similar to literal-based seeding. Thus, \codeIn{19} is added back in this case.

\paragraph{Value generation}
Finally, columns are generated in dependency order. Columns without constraints are populated by sampling from their per-column candidate pool uniformly until the assigned cardinality is reached. \codeIn{PRIMARY KEY} and \codeIn{UNIQUE} columns are populated first by sampling each value at most once from the pool; if the pool is exhausted before the required cardinality is reached, \tool{} synthesizes new unique values and extends the pool until all rows are filled. Foreign key columns are then populated by sampling from the already-generated values of the referenced column (a primary or unique key of the parent table), guaranteeing referential integrity. \tool{} also handles relational \codeIn{CHECK} constraints between two columns of the same table (\eg{} \codeIn{EMP.age > EMP.dept_id}) for each row. Specifically, when generating a column involved in such constraints, \tool{} checks whether the other column has already been generated. If not, the constraint is deferred, and the current column is sampled freely. Otherwise, the candidate pool is restricted to values that satisfy the constraint against the already-generated column's values, and the current column is sampled from the remaining candidates. Throughout all stages, multiple constraints on the same column are composed conjunctively, with each constraint further narrowing the candidate pool until only values satisfying all constraints remain. For \codeIn{UNIQUE} columns, we additionally check that any synthesized values also satisfy all other constraints on that column.\footnote{Termination is guaranteed, because the domain of each type is significantly larger than the target cardinality. However, if the combination of \codeIn{UNIQUE} and \codeIn{CHECK} constraints restricts the number of valid distinct values to less than the required cardinality, generation becomes infeasible.}

\subsection{Coverage-guided Feedback}\label{sec:coverage}
In this section, we describe the coverage-guided feedback mechanism in \tool{} (see Figure~\ref{fig:walkthrough} (\ding{175})). We use coverage to guide two aspects of fuzzing: (1) increasing the cardinality of generated tables when coverage stagnates to encourage the exploration of new behaviors, and (2) terminating the fuzzer once coverage has saturated, and further generation is unlikely to yield new outcomes. 



\paragraph{Approximate predicate coverage}


\begin{lstlisting}[style=sql, label={lst:cov}, caption={Approximate Predicate Coverage example}, float]
~-- Original query~
SELECT AVG(age) FROM EMP
JOIN DEPT ON @DEPT.id = EMP.dept_id@
WHERE @EMP.age >= 20@ AND @DEPT.code = 'CS'@;
~-- Coverage query~
SELECT @EMP.age>=20@, @DEPT.code='CS'@, @DEPT.id=EMP.dept_id@
FROM EMP CROSS JOIN DEPT;
\end{lstlisting}

We propose \emph{Approximate Predicate Coverage (APC)}, a coverage metric that acts as a \emph{proxy} for a query's execution paths by tracking its predicate outcomes. Since the evaluation of predicates determines how data is processed, tracking them effectively approximates the query logic~\cite{tuya10sqlfpc}. Consequently, APC heuristically indicates whether generated database instances could discover new query behaviors. Formally, let $P$ be the union of all $n$ distinct predicates extracted from both queries in the input pair, and $\tau(p, D) \subseteq \{\codeIn{TRUE}, \codeIn{FALSE}, \codeIn{NULL}\}$ denote the set of distinct ternary outcomes observed for a predicate $p \in P$ when evaluated against a database instance $D$. We denote the event of a predicate $p$ evaluating to a specific outcome $v$ as $p^v$. APC is defined as the set of all distinct predicate outcomes triggered across the entire instance:$$\text{APC}(D) = \{ p^v \mid p \in P, v \in \tau(p, D) \}$$Under this definition, the maximum possible coverage size is $3n$. A new database instance $D'$ increases coverage if $\text{APC}(D') \not\subseteq \text{APC}(D)$, indicating that at least one predicate reached a previously unseen ternary state.

To calculate APC, we systematically transform each query in the input pair into two distinct \emph{coverage queries}. Consider an input query $Q$ that references tables $T_1,\ldots,T_n$. Let $p_1,\ldots,p_k$ denote the conjuncts from \codeIn{ON} clauses, $p_{k+1},\ldots,p_m$ denote the conjuncts from the \codeIn{WHERE} clause, and $p_{m+1},\ldots,p_o$ denote the conjuncts from the \codeIn{HAVING} clause. APC maps $Q$ to two coverage queries, $Q_r$ and $Q_a$. The row-level coverage query, $Q_r$, selects the truth-values $\langle p_1\rangle,\ldots,\langle p_m\rangle$ from the cross product $T_1 \times \cdots \times T_n$. It therefore evaluates the \codeIn{ON} and \codeIn{WHERE} predicates over every row combination, with all referenced tables kept in scope. For example, in Listing~\ref{lst:cov}, we first extract the referenced tables (\codeIn{EMP} and \codeIn{DEPT}) along with all row-level predicates (highlighted in red). We then construct the coverage query by projecting these predicates over the cross-joined tables. If $Q$ contains the \codeIn{GROUP BY} clause, APC constructs the aggregate-level coverage query $Q_{a}$ by preserving the original clauses except \codeIn{WHERE} and \codeIn{HAVING}, and selects $\langle p_{m+1}\rangle,\ldots,\langle p_o\rangle$. It therefore evaluates the \codeIn{HAVING} predicates over groups from the unfiltered rows, so their truth values can be observed for coverage. Overall, APC covers a subset of predicates from the original query. Predicates nested within complex constructs, such as subqueries and window functions, as well as predicates involving dialect-specific features, are intentionally left out of scope to keep the approach scalable and DBMS-agnostic. Importantly, APC is used only for fuzzing feedback; omitting a predicate from APC preserves the full data generation space, although it may make the fuzzing guidance less informative.


For each generated database instance, \tool{} executes the coverage queries once against it to calculate APC. For example, consider the generated instance $D$ in Figure~\ref{fig:walkthrough}. Since both queries in the input pair contain the same tables and predicates, they share the same coverage query. Executing this coverage query in Listing~\ref{lst:cov} against $D$ returns two rows of evaluations, as shown in the ``Predicate coverage'' box. If we let $p_0$, $p_1$, and $p_2$ represent the three predicates from left to right, the first row evaluates to \codeIn{FALSE}, \codeIn{TRUE}, and \codeIn{TRUE}, while the second row evaluates to \codeIn{TRUE}, \codeIn{TRUE}, and \codeIn{NULL}. Applying our formal definition, the resulting coverage set is $\text{APC}(D) = \{p_0^F, p_0^T, p_1^T, p_2^T, p_2^{N}\}$. 

In cases where neither input query contains any predicates, the coverage set $\text{APC}(D)$ remains empty. Conceptually, a query with no predicates has no data-dependent logical boundaries to explore, making coverage-guided feedback unnecessary. As a result, the fuzzer defaults to its stagnation scheduling behavior since no new coverage can be observed.

\paragraph{Cardinality scheduling}
To balance execution overhead with the need to explore diverse predicate behaviors, \tool{} dynamically schedules the maximum table cardinality during data generation. Intuitively, generated tables should initially be small to allow for rapid execution. As the fuzzer progresses, it should generate progressively larger tables to explore more complex database states. Furthermore, if APC stagnates and fails to trigger new query behaviors, the tool must actively expand the search space by generating even more rows to uncover unexplored paths. A naive alternative approach might enumerate all input combinations in a single table. However, exhaustively combining all possible value assignments across multiple columns, rows, and tables leads to a combinatorial explosion. Generating instances to exhaustively cover this large search space is practically infeasible.


We implement this scheduling through an upper-bound adjustment. In each iteration, the target table cardinality is sampled uniformly from $[0, b]$. To prioritize fast exploration, the upper bound $b$ is initialized to 1 in the first iteration. The base value of $b$ increments by 1 in each subsequent iteration, allowing the fuzzer to explore progressively larger database instances. We apply an additional expansion factor when coverage stagnates. Specifically, whenever coverage stagnates, $b$ is incremented by an extra 1 for every consecutive iteration that fails to yield new coverage.




\paragraph{Early termination}
To estimate whether continuing to run the fuzzer is likely to discover new query behaviors, we propose an early termination criterion for \tool{} based on the \emph{Local Laplace Estimator (LLE)}. While the classical Laplace estimator is effective for estimating stoppage in black-box fuzzing, where the input space is fixed~\cite{bohme2018stads}, \tool{} is conceptually a greybox fuzzer that dynamically expands its input space by increasing table cardinality, changing the sampling distribution. Thus, based on prior theoretical and empirical results~\cite{bohme2018stads, estimatResidual21Bohme}, we adopt a reset-based estimator to handle this nonstationarity. Specifically, the classical Laplace estimator is defined as $(s+1)/(k+2)$, where $s$ is the number of successful discoveries among $k$ inputs~\cite{estimatResidual21Bohme}. Since no counterexample is found initially, $s = 0$, and
the estimator reduces to $1/(k+2)$. Whenever new query behaviors are observed, we reset the estimator to reflect the renewed potential for discovery within the expanded input space. Formally, we define the LLE as:$$\hat{p}(k) = \frac{1}{k + 2}$$where $k$ is the number of consecutive iterations \emph{without} observing new APC. Whenever new coverage is discovered, $k$ resets to 0. Fuzzing terminates when the estimated probability falls below a user-specified risk threshold, $\hat{p}(k) < \delta$. In other words, the fuzzer stops when no new query behaviors have been discovered for a sustained period. For example, if the fuzzer executes 8 consecutive iterations without new coverage, the estimated probability of the next iteration being productive is $\hat{p}(8) = 1/10 = 0.1$

\section{Implementation}

\sloppy{}
We implemented \tool{} in Python 3.11 in approximately 4,500 lines of code. We use SQLGlot~\cite{SQLGlot}, a multi-dialect parser, to extract literals and construct coverage queries. We use SQLAlchemy~\cite{sqlalchemy} as a DBMS-agnostic interface to issue \codeIn{CREATE} and  \codeIn{INSERT} statements uniformly across backends. We set $\delta = 0.05$ as the termination threshold. \tool{} supports five commonly used data types, shown in Table~\ref{tab:datatypes} along with their initial candidate pools and boundary functions. We map Python data types to their DBMS equivalents via SQLAlchemy. Supporting additional data types requires specifying only an initial pool $I$ and a boundary function $B(v)$.



\begin{table}[]
\centering
\footnotesize
\caption{Supported data types.}
\vspace{-4pt}
\label{tab:datatypes}
\begin{tabular}{lll}
\toprule
\textbf{Data Type} & \textbf{Initial Values $I$} & \textbf{Boundary Function $B(v)$} \\
\midrule
Integer   & $\{0,\ 1,\ \texttt{NULL}\}$                                    & $\{v-1,\ v,\ v+1\}$ \\
Float     & $\{0.0,\ 0.1,\ \texttt{NULL}\}$                               & $\{v-1.0,\ v,\ v+1.0\}$ \\
String    & $\{\texttt{default1},\ \texttt{default2},\ \texttt{NULL}\}$   & $\{v\}$ \\
Date/Time & $\{\texttt{2023-01-01},\ \texttt{2023-01-02},\ \texttt{NULL}\}$ & $\{v-1,\ v,\ v+1\}$ \\
Boolean   & $\{\texttt{TRUE},\ \texttt{FALSE},\ \texttt{NULL}\}$          & $\{v\}$ \\
\bottomrule
\end{tabular}
\end{table}
\section{Evaluation}\label{sec:evaluation}
In this section, we present a large-scale evaluation of \tool{} addressing the following questions.
\begin{itemize}[leftmargin=*]
\item \textbf{Q1:} How effectively does \tool{} refute equivalence (\ie{} how many non-equivalent query pairs can it successfully refute)?
\item \textbf{Q2:} How efficiently does \tool{} refute query equivalence (\ie{} how much time does it take)?
\item \textbf{Q3:} How do key components in \tool{}, such as the termination threshold and literal-based seeding, impact its performance?
\item \textbf{Q4:} Is \tool{} useful in downstream applications?
\end{itemize}

\subsection{Experimental Setup}
In this section, we describe the benchmark suites and baselines used in our evaluation. All subsequent experiments were run on a server with a 64-core AMD EPYC 7763 processor (2.45GHz) and 512GB of RAM, running Ubuntu 22.04.


\paragraph{Benchmarks}
We evaluated SQLWitness on benchmark suites containing both human-written and LLM-generated query pairs. Table~\ref{tab:benchmark} presents the benchmark suites used in our evaluation, with a total of 26,665 query pairs in four different dialects. Each benchmark suite links to a specific DBMS, as query pairs are written in a particular SQL dialect and are intended to be executed against the corresponding system. A key property of these benchmarks is that many query pairs are semantically inequivalent, making them well-suited for evaluating refutation tools. We retained only query pairs where both executed successfully. Including non-executable queries would bias the evaluation in favor of test-case-based tools---which trivially detect inequivalence via syntax errors raised by the DBMSs---and penalize verification-based tools---since they would need to identify semantic incorrectness through their model, which is more challenging. Note that we were unable to determine the number of \emph{equivalent} query pairs in the benchmarks, as the large scale of the benchmarks makes manual annotation infeasible, which was also previously observed~\cite{verieql24he, zhao25polygon, chen25Parseval}.


\begin{table}[]
\caption{Benchmark suites used in our evaluation. We use T-SQL to refer to the dialect used in the \sede{} benchmark.}
\centering
\label{tab:benchmark}
\small
\begin{tabular}{llllr l}
\toprule
\textbf{Benchmark} & $\mathbf{Q_1}$ \textbf{source} & $\mathbf{Q_2}$ \textbf{source} & \textbf{\#Pairs} & \textbf{DBMS} \\
\midrule
LeetCode \cite{verieql24he} & Solutions & Submissions & 22,778 & MySQL 8.2.0 \\
Spider \cite{spider1} & Dev-set  & OmniSQL \cite{li25Omni} & 1,032 & SQLite 3.50.2 \\
\bird{} \cite{liCanLLMAlready} & Dev-set  & OmniSQL \cite{li25Omni} & 1,515 & SQLite 3.50.2 \\
Mini-Dev$_M$ \cite{liCanLLMAlready} & Dev-set  & GPT-5.4 & 438 & MySQL 8.2.0 \\
Mini-Dev$_P$ \cite{liCanLLMAlready} & Dev-set  & GPT-5.4 & 407 & PostgreSQL 17.4 \\
\sede{} \cite{hazoom2021TexttoSQLWildNaturallyOccurring} & Test-set  & GPT-5.4 & 495 & SQL Server 2022 \\
\bottomrule
\end{tabular}
\end{table}

\paragraph{Baselines}
We compared \tool{} against the most relevant state-of-the-art query refutation tools that are publicly available. Specifically, we selected the following tools as baselines:
\begin{itemize}[leftmargin=*]
    \item \textbf{\parseval{}.} \parseval{} is a test-case-based query refutation tool which uses constraint solving to generate database instances that cover all branches of a query plan~\cite{chen25Parseval}. Similar to \tool{}, we expect \parseval{} to produce all genuine counterexamples since it validates generated instances on a database. However, due to the use of a constraint solver, we expect it to be less efficient than \tool{}. {We evaluated the most recent stable version of \parseval{} available at time of submission (commit \codeIn{5e80792}).}
    \item \textbf{\verieql{}.} \verieql{} is a bounded SQL equivalence verifier that can also disprove equivalence and generate counterexample witnesses~\cite{verieql24he}. We expect \verieql{} to produce no false positives under its defined semantics, but at the cost of efficiency at larger bounds. It also lacks support for certain SQL features, such as window functions and correlated subqueries.
    \item \textbf{SpotIt.} SpotIt extends \verieql{} to support the query features found in Text-to-SQL benchmarks such as \bird{}, including data types (\eg{} floats) and functions such as \codeIn{STRFTIME}~\cite{klopfenstein26spotit}. Despite this broader coverage, we expect SpotIt to share the same fundamental limitations as \verieql{}.
    \item \textbf{\polygon{}.} \polygon{} is a bounded symbolic reasoning engine for SQL that uses an under-approximation search algorithm and supports multiple applications, including query refutation~\cite{zhao25polygon}. We expect \polygon{} to be highly efficient due to its underlying algorithm, but to miss more inequivalent pairs or produce incorrect counterexamples due to its approximate semantics encoding. Since it shares the same semantics as \verieql{}, we expect both tools to support a roughly comparable set of queries.
\end{itemize}
We excluded several tools due to limitations that make them unsuitable for our evaluation settings. Specifically, QED~\cite{wang24QED} and Cosette~\cite{chu17cosette} lack foreign key support, making them incompatible with almost all of our benchmarks, as they contain such constraints. Without considering these constraints, many data instances would fail to be inserted into the databases~\cite{verieql24he}. We further excluded tools for which prior work reported significantly lower query support rates or refuted rates than the state-of-the-art~\cite{chen25Parseval, zhao25polygon, verieql24he}, including SQLSolver~\cite{sqlsolver}, XData~\cite{chandra2015DataGeneration}, and Spider-TS~\cite{2020spiderts}.


\subsection{Q1: Effectiveness}\label{sec:q1}
In this section, we study \tool{}'s effectiveness in refuting query equivalence, specifically, its ability to discover inequivalent query pairs and generate valid counterexamples for them. 


\paragraph{Metrics}
We evaluated each tool's effectiveness in terms of its ability to successfully refute inequivalent query pairs, while avoiding unsupported queries and spurious counterexamples. For each query pair, a tool produces one of four outcomes. A query pair is \emph{refuted} if a genuine counterexample was found; we report the \emph{Refuted Rate}, the fraction of refuted pairs, as our primary effectiveness metric (higher is better). A query pair is \emph{checked} if the tool determined the pair to be potentially equivalent, either via bounded equivalence or by terminating without finding a counterexample, following the terminology of \verieql{}~\cite{verieql24he}. A query pair is \emph{unsupported} if the tool failed to process it due to an error or timeout; we report the \emph{Unsupported Rate} as the fraction of such pairs (lower is better). Finally, the outcome of a query pair is \emph{spurious} if the produced counterexample returns the same result when executed on the DBMS, indicating an unsound witness; we report the \emph{Spurious Rate} as the fraction of such pairs (lower is better).

\paragraph{Methodology}
We evaluated all baselines and \tool{} on each benchmark. For each refuted query, we validated the found counterexample by re-running it on the DBMS, checking whether the counterexample instance satisfies integrity constraints and whether the query outputs genuinely differ. Counterexamples that passed this check were counted as genuine; those that failed were treated as spurious. For all tools, we set the timeout to 60 seconds, following \polygon{}'s experimental setup~\cite{zhao25polygon}. We chose bound 2 for the bounded tools (\verieql{}, SpotIt, and \polygon{}), as bound 1 found significantly fewer counterexamples (45\% fewer on Spider for \verieql{}) while bound 3 incurred substantially greater runtime (13.6$\times$ slower on \bird{} for \verieql{}) with only marginal gains. This sensitivity to the choice of bound also highlights that bounded tools require careful tuning to perform at their best. For \tool{}, we ran it ten times and report the mean, as the tool is non-deterministic.

\paragraph{Refuted rates}
\begin{figure}
    \centering
    \includegraphics[width=0.9\linewidth]{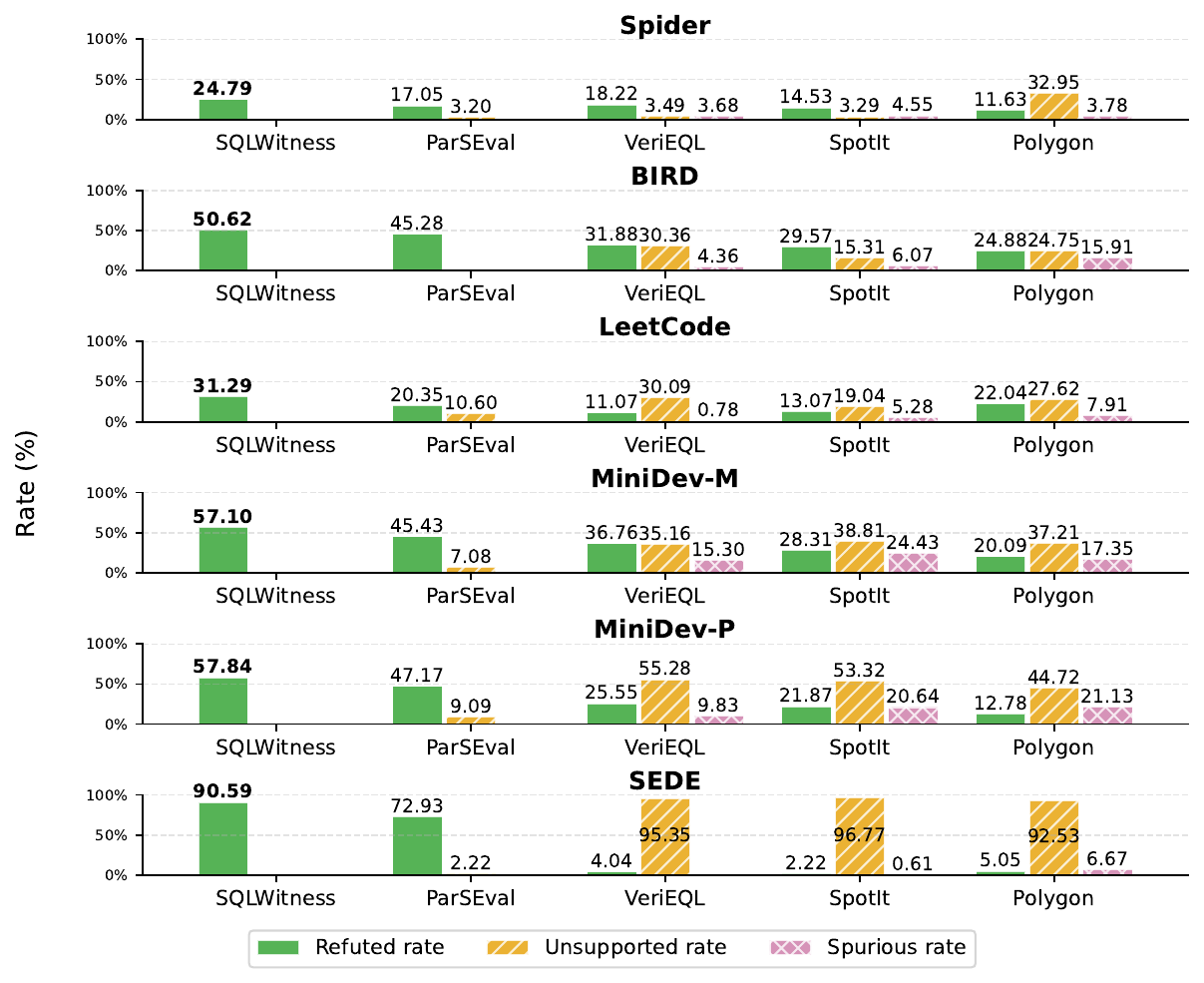}
    \caption{{Refuted, unsupported, and spurious query pair percentage. \tool{} achieved higher refuted rates than all baselines, with a standard deviation of less than 0.5\%.}}
    \label{fig:effectiveness}
\end{figure}


We present the results of \emph{Refuted Rates}, \emph{Unsupported Rates}, and \emph{Spurious Rates} in Figure~\ref{fig:effectiveness}. Across the six benchmarks, \tool{} refuted {5.34\%} to 88.37\% more query pairs than the baselines, demonstrating the effectiveness of the query-aware fuzzing strategy. The improvements in refuted rate were more significant on benchmarks with dialect-specific features, such as Mini-Dev and \sede{}, where \tool{} substantially outperformed all baselines, highlighting its DBMS-agnostic capability.


\paragraph{Unsupported rates}
\tool{} supported all SQL query pairs through its lightweight syntax-based analysis. The verification-based baselines consistently encountered at least 20\% of unsupported queries on most benchmarks, with the unsupported rate increasing substantially on benchmarks with more dialect-specific features, reaching at least 44\% on Mini-Dev$_P$ and 95\% on \sede{}. SpotIt extended \verieql{} to support SQLite-specific features, reflected in its lower unsupported rate on \bird{} compared to \verieql{}; however, the two tools showed only marginal differences on other dialects. Moreover, despite its broader feature support, SpotIt refuted slightly fewer query pairs than \verieql{}, suggesting possible overfitting to the newly added features. {\parseval{} had the lowest unsupported rates among the baselines, ranging from 0.40\% on Bird to 10.60\% on LeetCode. This is because it generates test data by solving constraints over query plans and delegates evaluation to the DBMS, and it fails mostly due to the solver exceeding the time budget. Notably, on \sede{}, \parseval{} supported 97.78\% of the query pairs while the other three baselines failed on the majority, which we attribute to their limited T-SQL parser support.} While it is feasible to extend existing tools by manually encoding the semantics of new features, the effort required to fully support even a single SQL dialect is substantial, making \tool{} a more practical and scalable solution.


\paragraph{Spurious rates}
We observed spurious counterexamples from the verification-based baselines, with spurious rates ranging from 0.61\% to 24.43\%. Spurious counterexamples resulted primarily from non-deterministic queries, where an \codeIn{ORDER BY} clause does not uniquely determine the sort order, and the DBMS must arbitrarily break ties. Listing~\ref{lst:spurious} shows one such case adapted from the benchmarks, with the differing parts highlighted in red. The two queries are logically equivalent, differing only in the order of the joined tables. However, when grouping by \codeIn{DEPT_ID} produces equal counts across groups, the DBMS must choose which group appears first under \codeIn{ORDER BY COUNT(*)}, and \codeIn{LIMIT 1} then returns whichever row that tie-breaking decision selects. Since the two join orderings can influence the physical row order, the DBMS may resolve the tie differently for each query, producing inconsistent results. This query pair was flagged as a spurious counterexample from \verieql{} at a bound of 2 when run on SQLite, where both queries returned \codeIn{-1}. 


Notably, the tie-breaking behavior varies across DBMSs. On MySQL, the counterexample is genuine, given the particular data insertion order in Listing~\ref{lst:spurious}. However, swapping the insertion order (\codeIn{DEPT VALUES (-1), (0)}) causes both queries to return \codeIn{-1}. On SQLite, our evaluation setting, both queries return \codeIn{-1} regardless of data insertion order. On other DBMSs such as ClickHouse and Firebird, both queries return the same result but tie-break to \codeIn{0} rather than \codeIn{-1}, further illustrating how non-deterministic behavior is DBMS-specific and difficult to model formally. Since \tool{} is execution-based, it captures the exact semantics of the target DBMS, making its results reliable and consistent with real-world behavior without requiring any explicit modeling of tie-breaking rules and subtle semantic differences.


\begin{lstlisting}[style=sql, float, caption={Non-deterministic query pair, diff highlighted}, label={lst:spurious}]
CREATE TABLE DEPT (ID INTEGER);
CREATE TABLE EMP (DEPT_ID INTEGER);
INSERT INTO DEPT VALUES (0), (-1);
INSERT INTO EMP VALUES (-1), (0);

SELECT E.DEPT_ID FROM @EMP@ AS E
JOIN @DEPT@ AS D ON @E.DEPT_ID = D.ID@
GROUP BY E.DEPT_ID ORDER BY COUNT(*) LIMIT 1;
~-- SQLite (-1); MySQL (0)~
SELECT E.DEPT_ID FROM @DEPT@ AS D
JOIN @EMP@ AS E ON @D.ID = E.DEPT_ID@
GROUP BY E.DEPT_ID ORDER BY COUNT(*) LIMIT 1;
~-- SQLite (-1); MySQL (-1)~
\end{lstlisting}

\begin{figure}
    \centering
    \includegraphics[width=0.75\linewidth]{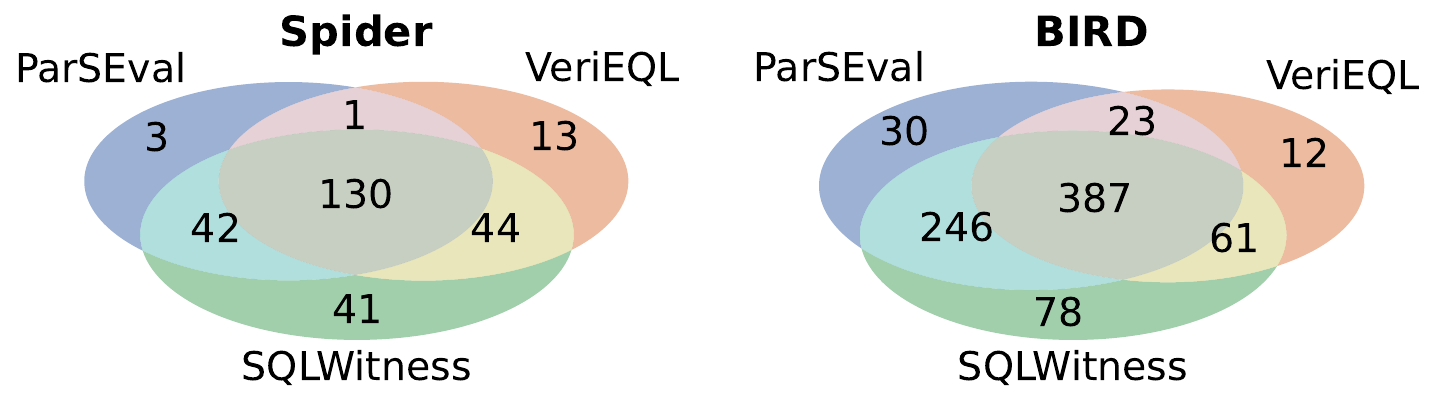}
    \caption{Venn diagrams of the overlap of refuted query pairs}
    \label{fig:overlap}
\end{figure}

\paragraph{Overlap analysis}
We analyze the overlap between the refuted query pairs of the verification-based baselines and \tool{} to investigate the potential benefit of combining both approaches. Figure~\ref{fig:overlap} presents Venn diagrams on the \bird{} and Spider benchmarks showing the overlap among the two best-performing baselines, \verieql{} and \parseval{}. {\tool{} uniquely refuted more query pairs than either baseline on both benchmarks (41 pairs on Spider and 78 on \bird{}), while also identifying 92.7\% and 91.4\% of the counterexamples found by the
baselines}, suggesting that most counterexamples discoverable by the baselines could be found by \tool{} alone. 



We sampled query pairs that \tool{} uniquely refuted but the baselines failed, identifying two main causes. The first is due to unsupported SQL features, as shown in Figure~\ref{fig:effectiveness}. When encountering unsupported features, the baseline tools explicitly emitted an error. The second and more noteworthy cause is semantic modeling gaps in how formal baselines encode query semantics relative to actual DBMS execution, which can cause them to treat inequivalent queries as equivalent. We noticed that string handling is particularly challenging, even without the presence of complex features such as the use of specific collations. For instance, \verieql{} silently considers a trivial pair as equivalent where one query filters with \codeIn{name = `a'} and the other with \codeIn{name = `a' OR name2 = `b'}, despite these being clearly inequivalent. \tool{} adds both string literals into the candidate pool and immediately generates a counterexample where \codeIn{name2 = `b'} is satisfied but \codeIn{name = `a'} is not. This issue also extends to more subtle cases where \codeIn{SUBSTR(date, 1, 4) = `2013'} and \codeIn{date BETWEEN `201301' AND `201312'} appear semantically equivalent but diverge on the string \codeIn{`2013'}, which satisfies the \codeIn{SUBSTR} predicate but falls below \codeIn{`201301'} under lexicographic comparison, excluding it from the \codeIn{BETWEEN} range. \tool{} adds \codeIn{`2013'} to the candidate pool via literal seeding and naturally generates this witness. Finally, when two queries differ in their \codeIn{GROUP BY} key, one grouping by a string column and the other by an integer column, \verieql{} again silently assumes equivalence. In practice, these findings suggest that formal tools should consider reporting an explicit unknown verdict rather than silently assuming equivalence, as a false equivalence verdict could be misleading to users.






We further inspected all query pairs that \tool{} failed to refute, but the baselines successfully refuted, identifying three causes. The most common is that the counterexample is reachable in principle, but requires a precise combination of data to trigger. Due to randomness, such data combinations might be missed within the termination threshold. While lowering the termination threshold (\ie{} running for a longer time) would increase the probability of generating such data, it would do so at the cost of efficiency. The second case involves non-deterministic queries whose results depend on the order in which rows were inserted into the database, as described above. The third and rarest case is when the candidate pool lacks the values necessary to trigger the counterexample. For example, the difference between \codeIn{ORDER BY ABS(LONGITUDE)} and \codeIn{ORDER BY LONGITUDE} only triggers with negative \codeIn{LONGITUDE} values, but since the candidate pool contains only non-negative values in our implementation, \codeIn{ABS} leaves its input unchanged in every generated instance, and the semantic difference is never exposed. One mitigation would be to design heuristics that augment the candidate pool based on the arithmetic operations present in the query, such as negating numeric literals when \codeIn{ABS} is detected. However, designing such heuristics for every possible operator would require significant manual effort. 

These findings suggest that \tool{} and tools that formally model query semantics are complementary. \tool{} finds most counterexamples, while formal tools refute a small set of cases that \tool{} misses, particularly those requiring highly specific arithmetic reasoning. 


\paragraph{Query feature analysis}

We quantitatively analyzed whether each tool could expose a \emph{query feature difference}. A feature difference was exposed if the tool could refute equivalence based on this feature. In other words, removing this feature would have made the outputs match on the witness. We analyzed all 1,111 query pairs in the union of the refuted sets from the overlap analysis, so all tools were compared over the same set of known-non-equivalent pairs. Given the large number of pairs, we used an LLM-in-the-loop approach to classify the features. We first minimized each counterexample by delta debugging~\cite{zeller02delta}, where we deleted rows one at a time, kept a deletion if the pair's output still differed, and stopped when every remaining row was necessary. Next, we followed the LLM-as-a-judge methodology with a panel of judges~\cite{gu2025surveyllmasajudge}, all using Claude Fable 5 to label the features. This approach mitigated potential LLM misclassifications through multi-agent consensus, and we manually verified any pairs that failed to reach consensus. Specifically, a labeler agent examined each pair's counterexample and assigned the exposed feature differences, or flagged the pair as non-deterministic. A panel of three judges then independently audited every label, and we accepted an assignment when at least two judges agreed. The panel confirmed the initial labels for 94.4\% of pairs, and we verified the remaining 5.6\% manually. To validate further, we randomly sampled 50 labeled query pairs and annotated them manually. The manual annotations agreed with the LLM-assigned labels on every pair. A Clopper-Pearson confidence interval on this sample showed that, with 95\% confidence, the true labeling error rate was below 5.8\%.

Table~\ref{tab:taxonomy} reported per-tool refutation rates for each feature. \tool{} achieved the highest refutation rate on every feature, and refuted over 90\% of pairs for most features, with type coercion having the lowest at 82.4\%. \parseval{} showed the lowest refutation rate on \codeIn{NULL} handling, as its coverage design failed to cover \codeIn{NULL} branches. \tool{}'s three-valued initialization covered \codeIn{NULL} values directly. \verieql{} had the lowest rate on string semantics and type coercion, reflecting the limitations of its formal encoding of such features discussed in the overlap analysis. We also examined the 16 refuted pairs due to non-determinism. Fifteen of the 16 pairs followed the \codeIn{ORDER BY} on non-unique columns pattern; the remaining one depended on a scalar subquery matching multiple rows, where the DBMS's arbitrary row choice decided the result. \verieql{} refuted 75\% of the pairs in this category, while notably, 54\% of \verieql{}'s spurious refutations were attributable to this same pattern.



\begin{table}[t]
\caption{{Refutation rates and number of query pairs per query
feature, over the union of pairs refuted by at least one tool. Each pair could contain multiple features.}}
\label{tab:taxonomy}
\small
\begin{tabular}{l r r r r}
\toprule
\textbf{Feature} & \textbf{\#Pairs} & \textbf{\tool{}} & \textbf{\parseval{}} & \textbf{\verieql{}} \\
\midrule
Join semantics   & 266 & \textbf{95.9\%} & 79.7\% & 58.6\% \\
Filter logic     & 204 & \textbf{91.2\%} & 85.3\% & 52.9\% \\
Aggregation      & 194 & \textbf{89.7\%} & 71.1\% & 59.8\% \\
Projection       & 187 & \textbf{95.7\%} & 94.7\% & 68.4\% \\
\codeIn{NULL} handling    & 180 & \textbf{96.7\%} & 60.6\% & 48.9\% \\
\codeIn{DISTINCT}         & 108 & \textbf{91.7\%} & 84.3\% & 88.9\% \\
String semantics &  51 & \textbf{88.2\%} & 76.5\% & 13.7\% \\
Ordering   &  43 & \textbf{97.7\%} & 55.8\% & 51.2\% \\
Set operation    &  21 & \textbf{95.2\%} & 71.4\% & 66.7\% \\
Type coercion    &  17 & \textbf{82.4\%} & 64.7\% & 11.8\% \\
\midrule
Non-deterministic &  16 & 31.2\% & 6.2\% & 75.0\% \\
\bottomrule
\end{tabular}
\end{table}

{
\paragraph{Counterexample pattern analysis}
We quantitatively analyze what \emph{data pattern} is needed to trigger each
counterexample, and whether \tool{} can reliably generate such data. Since no established or obvious metrics exist that could quantify the difficulty of finding equivalence counterexamples, we propose two approximate metrics that characterize the data pattern a counterexample requires. One captures constraints on individual values, and the other captures
constraints between values; we name them \emph{Pinned} and \emph{Aligned}. Pinned counts the data values that must pass or fail
a specific comparison in the query. Aligned counts the data values
that must be the same as other values. For example, if
one query has \codeIn{WHERE c = 0 OR c = 1 GROUP BY g HAVING COUNT(*)
> 3} and the other differs only in \codeIn{COUNT(*) >= 3}, a
counterexample needs exactly three rows that pass the \codeIn{WHERE}
clause and belong to one group. In this case, Pinned is 3 (each row's
\codeIn{c} must pass the \codeIn{WHERE} condition), and Aligned is 3
(the three rows' \codeIn{g} values must be the same). We compute the two metrics for every pair \tool{} refuted across ten runs on Spider and
\bird{}.




Table~\ref{tab:difficulty} shows the mean Pinned and Aligned counts,
grouped by how many of the ten \tool{} runs refuted the pair. The number of successful runs measures how reliably \tool{}
generates a counterexample. We observe that higher Pinned and Aligned values indicate less reliable runs, as more data must take a specific value or share the same value as each other, while randomly generated data is less likely to satisfy all such constraints at once. Most pairs (923 of 1{,}111) were refuted in every run, while the best baseline refuted only 79.5\%. Pairs that have higher Pinned and Aligned counts are more difficult for \tool{} to refute. For example, one pair that differs in \codeIn{COUNT(id)} versus
\codeIn{COUNT(DISTINCT uuid)} over a three-table join has a Pinned of 4
and an Aligned of 4. \tool{} refuted this pair in only one of the ten
runs. Such pairs need a precise data combination that random
generation may fail to produce reliably. However, for solver-based
tools, their refutation rates stayed roughly the same as the pairs
became more difficult for \tool{}.

We also examined why the 50 never-refuted pairs fail. For each pair, we took the baseline's refuting example where available, and calculated Pinned and Aligned values. The 50 pairs fall into the three failure types identified in the overlap analysis. First, 11 pairs are non-deterministic pairs that require a specific row insertion order to trigger. Second, 18 pairs require a data value that \tool{}'s candidate pool lacks. Third, the remaining 21 pairs require specific data combinations that our approach failed to generate, with a mean Pinned of 4.4 and a mean Aligned of 3.5.

\begin{table}[t]
\caption{{Refuted query pairs grouped by the number of \tool{}'s ten runs that refuted them, with mean Pinned and Aligned counts}}

\label{tab:difficulty}
\small
\setlength{\tabcolsep}{5pt}
\begin{tabular}{l r r r r r}
\toprule
\textbf{Refuted times} & \textbf{\#Pairs} & \textbf{Pinned} & \textbf{Aligned} & \textbf{\parseval{}} & \textbf{\verieql{}} \\
\midrule
10        & 923 & 1.3 & 2.2 & 79.5\% & 60.5\% \\
9        &  48 & 2.7 & 2.6 & 68.8\% & 58.3\% \\
1--8 & 90 & 3.1 & 4.1 & 74.4\% & 64.4\% \\
0      &  50 & --- & --- & 56.0\% & 54.0\% \\
\bottomrule
\end{tabular}
\end{table}
}

\subsection{Q2. Efficiency}

In this section, we investigate the efficiency of \tool{} in finding counterexamples.

\paragraph{Methodology}
We assessed the efficiency of \tool{} and the baselines on each benchmark by measuring the run time for each tool to produce an outcome for a query pair, which we term \emph{completing} a query pair. For each baseline, we excluded unsupported query pairs that failed immediately, as they contribute no meaningful insights. While this means that the tools are compared on different query pairs, we also evaluated efficiency on the common subset of {14,946} query pairs supported by all baselines, and observed the same efficiency ranking as reported below.

\paragraph{Efficiency}

\begin{figure}
    \centering
    \includegraphics[width=0.75\linewidth]{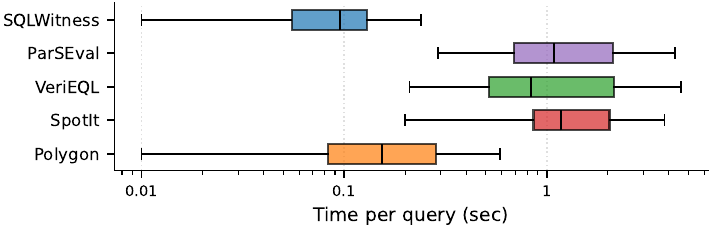}

    \caption{{Distribution of run time for all query pairs.}}
    \label{fig:efficiency_agg}
\end{figure}

Figure~\ref{fig:efficiency_agg} shows the distribution of per-query run time, and Table~\ref{tab:efficiency-mean} reports the arithmetic mean and the slowdown factor relative to \tool{}, computed as the geometric mean of per-query slowdown ratios over the pairs each tool could run. \tool{} achieved a mean of 0.12 seconds per query pair, outperforming all baselines. \tool{} also showed a median of 0.10 seconds, with most query pairs completing within a narrow range around its median, indicating consistent performance. Despite being test-case-based, \parseval{} was the slowest, with a 16.4$\times$ slowdown due to its use of constraint solvers, on par with SpotIt (16.1$\times$) and VeriEQL (13.3$\times$) at a bound of 2. \polygon{} was the second-most efficient tool after \tool{}, but it also produced the lowest refutation rate and the highest spurious rate in our evaluation. At bounds greater than 2, we believe \verieql{}, \polygon{} and SpotIt would be less efficient than the reported results. Overall, \tool{} is the most efficient and effective tool in our comparison.

The main efficiency bottleneck for \tool{} was coverage queries that reference multiple tables. Since coverage queries are constructed using \codeIn{CROSS JOIN}, with a large number of referenced tables, their execution speed could be slow when each table contains many rows. For instance, the slowest query pair in \bird{} involved six tables and eleven predicates, with \tool{} running until timeout in certain runs. This could be mitigated by restricting the size of the tables input to coverage queries, for example, by sampling rows from each referenced table before computing the cross product, trading completeness for efficiency. Still, since \tool{} begins by generating small instances, coverage queries over large tables are rarely triggered in practice.

\begin{table}[]
\centering
\small
\caption{Mean run time and slowdown for all query pairs.}
\begin{tabular}{l r r}
\toprule
\textbf{Tool} & \textbf{Mean (sec)} & \textbf{Slowdown} \\
\midrule
SQLWitness & 0.12 & --- \\
\parseval{} & 3.39 & 16.4$\times$ \\
\polygon{}  & 0.57 & 2.1$\times$ \\
VeriEQL    & 2.31 & 13.3$\times$ \\
SpotIt     & 2.28 & 16.1$\times$ \\
\bottomrule
\end{tabular}
\label{tab:efficiency-mean}
\end{table}

\subsection{Q3. Ablation study}
In this section, we investigate the contributions of important components of \tool{}.

\paragraph{Methodology} We designed two sets of ablations to assess the impact of the \emph{candidate pool} and \emph{termination threshold} components. For the candidate pool ablations, we compared three variants to examine their contribution to the effectiveness of refutation:
\begin{itemize}[leftmargin=*]
    \item \textbf{\tool{}$_{L}$} We removed the literal-based seeding mechanism, reducing the candidate pool to the three-value initialization set (\ie{} \codeIn{NULL} and two distinct concrete values).
    \item \textbf{\tool{}$_{LN}$} From \tool{}$_L$, we removed the \codeIn{NULL} value in all initial candidate pools (\ie{} all pools only contained two distinct concrete values).
    \item \textbf{\tool{}$_{L1}$} From \tool{}$_L$, we removed one of the concrete values in all candidate pools (\ie{} all pools only contained \codeIn{NULL} and one concrete value).
\end{itemize}

For the termination threshold ablations, we compared four variants denoted as \tool{}$_\delta$. Specifically, we varied the termination risk threshold $\delta$ in the termination mechanism, evaluating $\delta \in \{10\%, 5\%, 2.5\%, 1\%\}$, to assess its effect on refutation effectiveness and efficiency. For example, for $\delta=10\%$, we denote the configuration as \tool{}$_{10}$. \tool{}$_{5}$ is our default configuration.

We evaluated all variants on the benchmarks using the \emph{Refuted Rate}, and also report the mean time needed to complete all query pairs for the termination threshold ablations.

\paragraph{Component contributions}
\begin{table}[]
\centering
\small
\caption{Candidate pool ablation results in refuted rates. (SQLW: \tool{})}
\label{tab:ablation}
\begin{tabular}{l ccc c}
\toprule
\textbf{Dataset}
  & SQLW$_{L}$
  & SQLW$_{LN}$
  & SQLW$_{L1}$
  & SQLWitness \\
\midrule
Spider             & 17.25\% & 16.76\% & 15.60\% & \textbf{24.79\%} \\
\bird{}      & 16.04\% & 14.98\% & 15.18\% & \textbf{50.62\%} \\
LeetCode           & 28.37\% & 11.49\% & 25.69\% & \textbf{31.29\%} \\
Mini-Dev$_M$        & 16.21\% & 15.98\% & 15.53\% & \textbf{59.36\%} \\
Mini-Dev$_P$        & 17.44\% & 15.97\% & 16.46\% & \textbf{57.25\%} \\
\sede{}               & 40.61\% & 36.97\% & 24.44\% & \textbf{90.51\%} \\
\midrule
Mean Drop & 50.90\%{\color{red}$\downarrow$} & 61.74\%{\color{red}$\downarrow$} & 57.18\%{\color{red}$\downarrow$} & --- \\
\bottomrule
\end{tabular}
\end{table}
\begin{figure}
    \centering
    \includegraphics[width=0.7\linewidth]{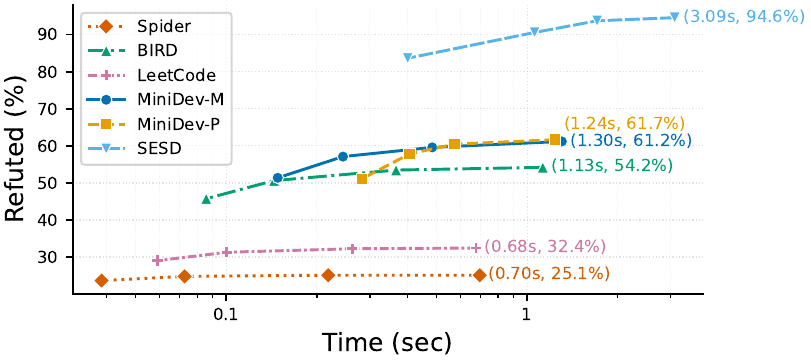}
    \caption{Termination threshold ablation results on refuted rate and run time. Each point on a curve corresponds to a value of $\delta \in \{10\%, 5\%, 2.5\%, 1\%\}$ from left to right.}
    \label{fig:threshold_ablation}
    
\end{figure}

Table~\ref{tab:ablation} shows that removing any component from the candidate pools lowers \tool{}'s performance. Removing literal-based seeding (\tool{}$_L$) yielded a geometric mean drop of 50.90\% in refuted rate across all benchmarks, indicating that literal values are critical for triggering query predicates and discovering counterexamples. Further removing one concrete value (\tool{}$_{L1}$) or \codeIn{NULL} (\tool{}$_{LN}$) showed larger drops of 61.74\% and 57.18\% respectively. These results demonstrate the effectiveness of the three-value initialization design. Notably, even with the initial candidate pool, \tool{}$_L$ already achieved refuted rates comparable to the baseline tools.

Figure~\ref{fig:threshold_ablation} presents the results of the termination threshold ablation, illustrating the effect of varying $\delta$ on both refuted rate and run time. We observed a clear trade-off between efficiency and effectiveness. Specifically, a lower threshold allows \tool{} to find more counterexamples at the cost of longer run times, with diminishing returns as the threshold decreases. For instance, reducing to $\delta=1\%$ yielded the highest refuted rate across all benchmarks, with an increase over \tool{}$_{5}$ at only around 2\%, while the mean run time increased by roughly 1 second per query pair, which was a 10$\times$ slowdown. Users could adjust $\delta$ to trade off between faster exploration and more exhaustive search depending on their needs, with $\delta=5\%$ constituting a balanced baseline.

\subsection{Q4. Use Case}\label{sec:usecase}
In this section, we demonstrate the applicability of our tool with a case study on testing dialect translation benchmarks. SQL dialect translators convert an SQL query written for one DBMS (\eg{} SQLite) into an equivalent query for another (\eg{} MySQL), which is crucial for applications such as database migration~\cite{zmigrodTranslatingSQLDialects2024}. Benchmarks such as \parrot{}~\cite{zhou2025parrot} and DLBench~\cite{li25dlbench} evaluate dialect translation tools using query pairs consisting of a source-dialect query and its target-dialect equivalent. We aimed to assess whether the reference translations of such benchmarks are accurate.

\paragraph{Methodology}
We used \tool{} to test an existing benchmark, \parrot{}~\cite{zhou2025parrot}. In \parrot{}, we selected pairs from the \bird{} benchmark, where the source dialect is SQLite and the target dialect is either PostgreSQL or MySQL. We collected all 379 SQLite-to-MySQL pairs and 488 SQLite-to-PostgreSQL pairs. We ran \tool{} with its default configuration on all pairs. To execute each query, we inserted the corresponding data into the respective DBMS, ran both queries, and compared their results.

\paragraph{Result Analysis}

Table~\ref{tab:parrot-analysis} presents a quantitative analysis of the counterexamples found among the \parrot{} dialect translations, categorized into two root causes. For the SQLite-to-MySQL translation, 76 query pairs were refuted, and for SQLite-to-PostgreSQL, 168 were refuted. We manually examined all refuted cases and identified two root causes. The first is \emph{Dialectal Difference}, where the translated query preserves the original logic, but the two DBMSs produce different results due to semantic differences between dialects, such as type conversion behaviors. The second is \emph{Execution Error}, where the translated query fails to execute on the target DBMS, typically due to syntax or run-time errors. For MySQL translations, the majority of counterexamples were Dialectal Differences (57.9\% of refuted pairs), while for PostgreSQL, the majority were Execution Errors (82.7\% of refuted pairs). This contrast highlights that PostgreSQL's strict syntax and type system cause more immediate execution failures, whereas MySQL's more permissive execution environment often masks deeper semantic divergences.



We present one example of each failure type. For MySQL, most Dialectal Differences were caused by implicit type conversion behaviors that differ from SQLite's, as illustrated in Listing~\ref{lst:avg} with the \codeIn{/} division operator. During translation, the division operator is preserved as-is, which produces differing results due to dialectal differences. To fix this, MySQL's \codeIn{DIV} operator should be used instead, as it performs integer division semantically equivalent to SQLite's \codeIn{/}. The Execution Errors for MySQL were mostly syntax errors; for instance, MySQL requires that any item in an \codeIn{ORDER BY} clause also appear in the \codeIn{SELECT} list, and certain unsupported functions were carried over directly from SQLite without translation, such as SQLite's null-safe summation function \codeIn{TOTAL}, which should be translated to \codeIn{COALESCE(SUM(), 0.0)}. PostgreSQL rejected many translations that followed SQLite's syntax conventions. For example, while SQLite permits arithmetic operations between integer and text types by implicitly coercing the text to a number, PostgreSQL throws an error and requires explicit casting. Due to these stricter guarantees, PostgreSQL translations produced fewer Dialectal Differences, where nearly all of them involve non-deterministic results or the same integer division discrepancy observed in MySQL.

These findings highlight the inherent difficulty of dialect translation and the shared challenges it poses for both translation tools and evaluation benchmarks. Correct dialect translation requires more than mapping unsupported syntax to equivalent supported features in the target dialect, as features that are supported across dialects can still carry different semantics, which might eventually lead to the translated queries failing to execute or returning incorrect results. Yet, existing benchmarks largely overlook this dimension, focusing on syntactic mapping correctness while failing to account for cross-dialect semantic differences. 






\begin{table}[]                                          
  \centering
  \small
  \caption{Analysis of refuted query pairs in \parrot{}.}                    
  \label{tab:parrot-analysis}                                                     
  \begin{tabular}{lrrr}                                                           
  \toprule
  \textbf{Translation} & \textbf{Refuted} & \textbf{Execution Error} & \textbf{Dialectal Difference} \\                                       
  \midrule
  MySQL      & 76 &  32 (42.1\%) & 44 (57.9\%) \\   
  PostgreSQL & 168 & 139 (82.7\%) &  29 (17.3\%) \\   
  \bottomrule                                                                               
  \end{tabular}                                                                             
\end{table}

\section{Discussion}\label{sec:discussion}

\paragraph{Effectiveness}
\tool{}'s heuristic approach reflects a fundamental tradeoff between fuzzing and
symbolic techniques, similar to challenges observed in software fuzzing~\cite{wu22onefuzzing}.
Symbolic tools like \verieql{} can precisely solve constraints to cover
arithmetic boundaries, but require formal encodings of operator semantics and struggle
with DBMS-specific features. Our evaluation shows that \tool{} already finds the vast
majority of counterexamples, yet some inequivalent pairs remain out of reach. The most
common failure mode is when a counterexample requires a precise arithmetic relationship
not captured by our candidate pool, such as negative values needed to distinguish
\codeIn{ABS(x)} from \codeIn{x}. Two directions can address this. First, \tool{} can
serve as a lightweight first stage in a pipeline that invokes a heavyweight formal
verifier only on pairs \tool{} fails to refute, combining fast practical coverage with
deeper formal reasoning. Second, LLMs could be used to generate operator-specific mutation
heuristics for arithmetic operators, augmenting the candidate pool while preserving
\tool{}'s DBMS-agnostic design.


\paragraph{Scalability}
\tool{}'s design is
DBMS-agnostic, and extending it to a new system requires minimal manual
effort. Specifically, a user needs to only implement an interface (\eg{} using
SQLAlchemy) for executing SQL statements (\eg{} \codeIn{CREATE} and \codeIn{INSERT}) on
the target DBMS, and for new data types, specify initial seed values and an analysis
pattern to extract literals of that type. Concretely, extending \tool{} to support
PostgreSQL required 28 lines of code, and adding support for date types required 34
lines of code. 




\section{Related Work}\label{sec:related}
\paragraph{Query equivalence verification}
Formal approaches have been the dominant strategy for SQL equivalence verification. Cosette~\cite{chu17cosette} and HoTTSQL~\cite{chu17hottsql} encode SQL semantics using U-semirings and univalent type theory to prove query equivalence. EQUITAS~\cite{zhou19equitas} and SPES~\cite{zhou22spes} translate queries into SMT formulae, with SPES extending support to bag semantics. QED~\cite{wang24QED} introduces Q-Expressions to model a larger class of query semantics while improving efficiency, and SQLSolver~\cite{sqlsolver} uses linear integer arithmetic to scale to additional query features. Bounded verifiers such as VeriEQL~\cite{verieql24he} and \polygon{}~\cite{zhao25polygon} support many SQL features and integrity constraints up to a fixed row bound. SpotIt~\cite{klopfenstein26spotit} extends VeriEQL with features like datetime functions to better support Text-to-SQL evaluation. All of these approaches rely on manual semantic encodings and lack support for the multitude of SQL dialects. \tool{} instead delegates all semantic interpretation to the underlying DBMS through execution, supporting arbitrary SQL features and DBMS-specific behaviors without any manual encoding effort.

\paragraph{Query behavior testing}
Various approaches generate tests to exercise query behavior. SQLFpc~\cite{tuya10sqlfpc} provides coverage criteria that measure how well test data exercises a query. XData~\cite{chandra2015DataGeneration} generates test databases that ``kill'' SQL mutants by solving symbolic constraints derived from query structure. Similarly, Spider-TS uses fuzzing to ``kill'' syntactically derived mutants to improve Text-to-SQL benchmark suites, but lacks coverage feedback. EvoSQL~\cite{castelin2018searchbased} uses a genetic algorithm to generate data achieving a target coverage criterion over a query. ParSEval~\cite{chen25Parseval} generates test databases covering all branches of a query's logical plan via symbolic execution, directly targeting equivalence refutation. While \tool{} is also test-case-based, it combines lightweight syntactic analysis with coverage-guided feedback for efficient query refutation.

\paragraph{SQL dialects}
SQL dialects represent a practical challenge in SQL tooling. In syntax, SQLFlex~\cite{an26sqlflex} leverages an LLM to achieve dialect-agnostic SQL parsing without additional manual effort. In semantics, Ye et al.~\cite{elucidatingType25Ye} studied type conversion behaviors across SQL engines, revealing inconsistencies in formal modeling. CrackSQL~\cite{zhou2025cracking} combines rule-based and LLM-based methods to translate queries across DBMSs, while Mallet~\cite{ngom2024mallet} uses LLMs with retrieval-augmented generation to produce reusable translation rules. Benchmarks such as \parrot{}~\cite{zhou2025parrot} and DLBench~\cite{li25dlbench} provide dialect translation pairs for evaluating dialect translation tools. \tool{} addresses the problem of finding concrete database instances that prove two queries inequivalent within one or across dialects, without requiring any dialect-specific semantic modeling.

\paragraph{Fuzz testing}
Fuzz testing is a widely adopted dynamic analysis technique for finding bugs in complex systems. AFL++~\cite{AFLplusplus-Woot20} is among the most widely adopted coverage-guided greybox fuzzers, using coverage feedback to guide mutation-based input generation. Fuzz testing has also been applied to testing DBMSs. SQLancer~\cite{Rigger2020TLP, Rigger2020NoREC} generates random queries and rewrites them into semantically equivalent queries to detect logic bugs, while SQLancer++~\cite{zhong2025scalingautomateddatabasetesting} provides a unified SQL generator that is applicable to most SQL-based DBMSs. Böhme's STADS framework~\cite{bohme2018stads} studies the stoppage criteria for black-box fuzzing by framing fuzzing as species discovery. Böhme et al. further explore residual risk in greybox fuzzing~\cite{estimatResidual21Bohme}, showing that classical estimators for black-box fuzzing fail as the input corpus shifts. \tool{} builds on these foundations, adapting the reset-based Laplace estimator for early termination. Wu et al.~\cite{wu25Tumbling} show that literal-based seed augmentation improves the effectiveness of greybox fuzzing, motivating \tool{}'s literal-based seeding heuristic. \tool{} adapts these foundations to SQL equivalence refutation, using approximate predicate coverage as a query-specific feedback signal and cardinality scheduling as the mechanism for exploring new predicate behaviors.
\section{Conclusion}
Existing query refutation tools often require significant manual effort and expertise to support new query features within and across dialects. In this paper, we have proposed \tool{}, an efficient DBMS-agnostic query equivalence refutation engine. It consists of a novel query-aware fuzzing approach that constructs minimal candidate pools to explore query behavior, generates constraint-aware data to ensure valid instances, and leverages predicate coverage with a statistical estimator to terminate the fuzzer. Our evaluation has shown that \tool{} is highly effective and efficient in refutation, having refuted 5.34\% to 88.4\% more pairs while achieving a 2.1$\times$ to 16.4$\times$ speedup compared to state-of-the-art approaches. We believe \tool{} could be adopted for many applications, which we demonstrated through the dialect translation use case, and requires minimal manual adaptation for new DBMSs. 

\begin{acks}
We would like to thank the anonymous reviewers for their constructive comments and suggestions. This project is supported by the Ministry of Education, Singapore, under the Academic Research Fund Tier 1 (FY2023).
\end{acks}

\clearpage
\bibliographystyle{ACM-Reference-Format}
\bibliography{ref}

\received{17 April 2026}
\received[revised]{19 August 2026}
\received[accepted]{12 September 2026}

\end{document}